\documentclass[aps,prd,twocolumn,
showpacs,
floatfix,nofootinbib,
superscriptaddress]{revtex4}

\usepackage{pstricks,epsfig,color}
\usepackage{psfrag}
\usepackage{epsfig}
\usepackage{epstopdf}

\newcommand{\be}{\begin{equation}}
\newcommand{\ee}{\end{equation}}
\newcommand{\bear}{\begin{eqnarray}}
\newcommand{\eear}{\end{eqnarray}} \newcommand{\ba}{\begin{array}}
\newcommand{\ea}{\end{array}}
\newcommand{\lae}{\begin{array}{c}\,\sim\vspace{-1.7em}\\<
\end{array}}
\newcommand{\gae}{\begin{array}{c}\,\sim\vspace{-1.7em}\\>
\end{array}}

\def\beq{\begin{equation}}
\def\eeq#1{\label{#1}\end{equation}}
\def\eeqn{\end{equation}}
\def\eeq{\end{equation}}
\def\beqa{\begin{eqnarray}}
\def\eeqa#1{\label{#1}\end{eqnarray}}
\def\eeqan{\end{eqnarray}}

\def\to{\rightarrow}

\newcommand\iden{\leavevmode\hbox{\small1\normalsize\kern-.33em1}}

\def\W3{W_H^3}

\begin{document}

\title{Resonances from Quiver Theories at the LHC}
\author{Gustavo Burdman
}
\author{
Nayara Fonseca
}
\author{Gabriela L.~Lichtenstein
}
\affiliation{Instituto de F\'isica, Universidade de S\~ao Paulo,
S\~ao Paulo SP 05508-900, Brazil}
\pacs{11.10.Kk, 12.60.-i, 13.90.+i}
\vspace*{0.3cm}


\begin{abstract}
We consider the collider signals of  spin-one resonances present  in full-hierarchy quiver theories of electroweak symmetry breaking.
These four-dimensional theories result from the deconstruction of warped extra
dimensional models and have very distinct phenomenological features
when the number of sites is small. We study a class of generic
scenarios in these theories where the color gauge group as well as the
electroweak sector, propagate in the quiver diagram. These scenarios
correspond to various specific models of electroweak symmetry breaking
and fermion masses. We focus on the minimum resonant content and its
main features: the presence of heavy and narrow spin one resonances. 
We derive bounds from the LHC data
on the color-octet and color-singlet excited gauge bosons from their
decays to jets and top pairs, and show their dependence on the number
of sites in the quiver.  We also  compare them with the bounds derived
from flavor violation.  

\end{abstract}

\maketitle

\section{Introduction}
\label{intro}
The electroweak standard model (SM) describes satisfactorily all
available data to date~\cite{pdg}. Since it is a renormalizable theory, this
implies that its cutoff $\Lambda$ --the scale of new physics--
is far above the weak scale $v\simeq246~$GeV. This has been  most recently
confirmed by the apparent discovery of a light Higgs boson with
$m_h\simeq 126~$GeV~\cite{higgsdicovery}, which is compatible with the
renormalizable SM Higgs sector.
On the other hand, the resolution of the hierarchy problem requires
that new physics beyond the SM appear at scales not too far above the
TeV. This little hierarchy problem points to the need to have a
parametric separation of the weak scale and the new physics scale.
In non-supersymmetric theories the Higgs must be a
remnant  pseudo-Nambu Goldstone boson (pNGB) from the spontaneous breaking of a
global symmetry~\cite{chm}.  The resonances will then have higher
masses as dictated by the gap between the pNGBs and the resonant
sector in analogy with the $m_\pi-m_\rho$ mass gap.
There are several scenarios beyond the SM with a pNGB Higgs.
These include the Little Higgs~\cite{littlehiggs}, Twin
Higgs~\cite{twinhiggs}, as well as extra-dimensional models where the
Higgs is obtained from a bulk gauge field in what is sometimes called
Gauge-Higgs Unification, particularly in AdS$_5$ backgrounds~\cite{ghunification}.
In all cases, there will be a large global symmetry spontaneously
broken giving rise to NGBs. Part of this global symmetry is gauged and
therefore explicitly broken.  This allows for a partial Higgs
mechanism eliminating some of the NGBs to give masses to the gauge
bosons associated with broken generators, and at the same time leads
to a potential for the Higgs and its small mass. For the model to be
successful, there must be a set of NGBs left out in the spectrum
forming a doublet of $SU(2)_L$ that can be identified with the
Higgs field responsible for EWSB.

The gap between $m_h$ and the resonant masses is a generic feature of
all these scenarios.  The tell-tale of the details of the underlying
dynamics  is in the resonant spectrum and couplings. It is possible to
parametrize these dynamics in an effective field theory framework of the low energy symmetries of
the SM. This has been done in several papers~\cite{silh}.

In this paper we will commit to a more specific set of models
including a pNGB Higgs. These theories can be represented by quiver (or moose)
diagrams~\cite{quiver1,quiver2} (see next section), and are cousins of
AdS$_5$ models since there is a limit in which the two are essentially
identical. In these limit, the quiver theories are obtained from the
deconstruction~\cite{decon,bbh} of AdS$_5$ theories. However, far form
this continuum limit, in what we can call the coarse limit, quiver
theories are four-dimensional and quantitatively very different from the
AdS$_5$ ones. In particular, the spectrum and couplings of the
resonant states --both bosonic and fermionic-- will be significantly
different than for the continuum case, and in general dependent upon
the number of gauge groups (or ``sites'' in the quiver diagram), as
well as the group structure and matter representation chosen.
Then, in the coarse deconstruction limit, quiver theories will have a
very distinctive phenomenology at the LHC.
We will begin exploring this phenomenology in vanilla quiver models as
the ones presented in Refs.~\cite{quiver1} and \cite{quiver2}. We will
concentrate on the production of vector resonances decaying into
quarks giving jets and $t\bar t $ pairs, 
as this should be the first signal for these models at the LHC
(as we show below).

The phenomenology of quiver or moose theories has been studied in many
other papers, but in different setups. For instance, in
Ref. ~\cite{higgsless3site} a three-site electroweak model without a
Higgs is built, and its phenomenology is studied in
~\cite{higgslesspheno}.  Its generalization to allow for a light Higgs
is presented in Ref.~\cite{hthreesite}. This ``221''  model is a very
specific quiver theory, and although there are quite a few common
points with our work, we will always consider larger gauge groups a a
set of ordered vacua. In Ref.~\cite{fdcompositeh} a two-site quiver is
proposed, its phenomenology of the extended gauge sector studied in
Ref.~\cite{fdcpheno}. A three-site construction more similar to ours is
that of Ref.~\cite{dischm}. Our approach differs from all these
previous contributions in one way or the other already at the model
building stage. We are considering generic coarse deconstruction
models with a very high ultra-violet cutoff. Our studies allow to
consider the number os sites as a variable. Our aim is to start a
systematic study of the phenomenology of quiver theories by pointing
out their main common features: narrow resonances as a result of
weak coupling, compatibility with flavor physics resulting in specific
decay channels, and a Higgs sector compatible with a pNGB light
Higgs. It is possible that some of our results can be partially applied to the
models mentioned above.

In the next section, we present the general framework for  quiver
theories, and we specify one model to study its phenomenology.
In Section~\ref{resonances}, we obtain the couplings of vector
resonances to SM fields, and in particular to SM quarks. We also
obtain the resonance widths. These results are used in
Section~\ref{bounds} to extract the current bounds on the model
spectrum from di-jet and $t\bar t$ resonance searches at ATLAS and CMS.
We give our conclusions and outlook  in Section~\ref{conclusions}.

\section{The Model Framework}
\label{model}
\begin{figure}
\begin{center}
\includegraphics[scale=0.56]{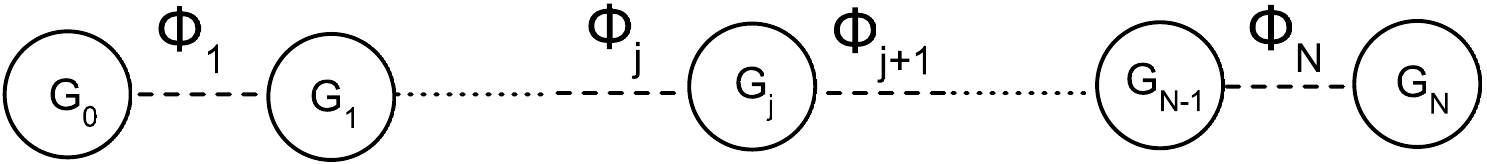}
\caption{Quiver diagram for the theory described by Eq.~(\ref{s1}).}
\label{f:1}
\end{center}
\end{figure}

We begin this  section by reviewing the basics of quiver theories (QT). Let us consider the product
gauge group $G_0\times G_1\times\cdots G_j\times G_{j+1}\cdots G_N$.
In addition, we have a set of scalar link fields $\Phi_j$, with
$j=1~{\rm to~} N$, transforming as bi-fundamentals under $G_{j-1}\times G_{j}$.
The action for the theory is
\bear
S &=& \int d^4x \left\{ -\,\sum_{j=0}^N\,\frac{1}{2g_j^2} \,Tr\left[ F_{\mu\nu}^{(j)}
    F^{\mu\nu (j)}\right]  \right.\nonumber\\
&&\left.+ \sum_{j=1}^N \,Tr\left[
    (D_\mu\Phi_j)^\dagger D^{\mu} \Phi_j\right] -V(\Phi_j)
+\dots
\right\}
\label{s1}
\eear
where the traces are over the groups'  generators, and the dots at the
end correspond to terms involving fermions and  will be discussed
in the next section. We assume that the potentials
for the link fields give each of them a vacuum expectation value (VEV) which breaks $G_{j-1} \times G_j$ down to the
diagonal group, and result in non-linear sigma models for the
$\Phi$'s
\be
\Phi_j = \frac{v_j}{\sqrt{2}}\, e^{i\sqrt{2} \pi_j^a \hat t^a/v_j}~,
\label{phis}
\ee
 where the $\hat t^a$'s are the broken generators, the $\pi_j^a$ the
 Nambu-Goldstone Bosons (NGB); and $v_j$ are the VEVs of the link
 fields. We will consider here the situation where the VEVs are
 ordered in such a way that $v_1>v_2\dots  > v_j\dots > v_N$.
We parametrize the ordering by defining the VEVs as
\be
v_j \equiv v q^j~,
\label{vjdef}
\ee
where $0<q<1$ is a dimensionless constant, and $v$ is a UV mass scale
that can be regarded as the UV cutoff.  We will also assume for the
moment that 
all the gauge groups are identical and that their
gauge couplings satisfy
\be
g_0(v)=g_1(v_1)=\dots=g_j(v_j) = g_{j+1}(v_{j+1})=\dots \equiv g~.
\label{gaugeequal}
\ee
The model can be illustrated
by the quiver diagram of Figure~\ref{f:1}.
The gauge boson mass matrix squared is given by
\begin{widetext}
\bear
M^2_g\,=\,g^2\,v^2\,\left(  
\renewcommand{\arraystretch}{1.6}
\begin{array}{ccccccc}
q^2   &  -q^2   & 0               &   0   & \cdots    & 0   &    0     \\
-q^2  &  q^2 + q^4  &   -q^4     &  0  & \cdots  &   0    &0      \\
0    & -q^4     &  q^4 + q^6    &  - q^6  & \cdots & 0 & 0   \\
\vdots & \vdots & \vdots  & \vdots &  \cdots  & \vdots& \vdots   \\
0    & 0 & 0 &0 &\cdots &    q^{2(N-1)} + q^{2N}    &  -q^{2N}  \\
0 & 0 & 0&0 &\cdots &      -q^{2N}     & q^{2N} 
\end{array}
\right)\,.
\eear

\end{widetext}
\noindent
in the basis $(A_0,A_1, \cdots,  A_N)$, and in the  unitary gauge.  We
diagonalize $M_g$ by the orthonormal rotation 
\be
A^j_\mu = \sum_{n=0}^{N} f_{j,n} A^{(n)}_\mu~,
\label{gaugerotation}
\ee
where the $\{A_\mu^{(n)}\}$ are the mass eigenstates.
The zero-mode gauge boson, $A_\mu^{(0)}$, has a ``flat profile'' in
the quiver diagram, meaning that for all $j$ $f_{j,0} =
1/\sqrt{N+1}$. This is not the case for the massive modes, for which 
$f_{j,n}$ can be obtained from the diagonalization procedure.  
In order to address the hierarchy problem, we will need that the first
gauge excitation is $v_N\simeq O(1)~$TeV. Furthermore, if we impose that these models are to
address the full hierarchy between the Planck and the electroweak
scales, then $v\lae M_{P}$. Thus, the values of the model parameter $q$
and the number of gauge groups $N$ would  be related by 
\be
q = 10^{-16/N}~.
\label{qvsN}
\ee
The Higgs field will have to have a profile highly localized towards
the site $N$, in order for the corrections to its mass to be no larger
than of  order of the electroweak  scale. In a full model this can be
done dynamically by extracting the Higgs doublet from a
NGB that stays in the spectrum~\cite{quiver2}. 
Here, we will make the simplification of assuming that the Higgs
doublet only transforms under the weak gauge group of site $N$,
i.e. is completely ``localized'' on the site $N$. This simplifying
assumption will be of little impact on the rest of the paper. 

In the limit of large $N$, and $q\to 1_-$, these models can be
described by the deconstruction~\cite{decon1} of theories with one 
compact extra dimension in an AdS background, AdS$_5$~\cite{rs}.
The deconstruction of AdS$_5$ was studied in
Refs.~\cite{decon1,bbh,tools}.   
This continuum limit, in which the four-dimensional theory described
above and the AdS$_5$ theories are equivalent, is  obtained when the
ultra-violet (UV) scale of the 4D theory, which is approximately $v$, is larger than the   
curvature $k$ of the 5D AdS space: $k<v$.  In fact, in the language of
the deconstructed theory obtained by discretizing AdS$_5$, $v$ corresponds
to the inverse of the discretization interval $a$. Using
Eq.~(\ref{qvsN}) and the identification $q\leftrightarrow e^{-k/gv}$~\cite{decon,bbh,quiver1}
necessary for matching both theories, we see that for $N\gae 36$ the
quiver theories would be essentially identical to the
extra-dimensional  theory in AdS$_5$. On the other hand, for smaller
values of $N$ the 4D theories cannot be interpreted as AdS$_5$ ones
and should be studied separately.  

\begin{figure}
\begin{center}
\includegraphics[scale=0.56]{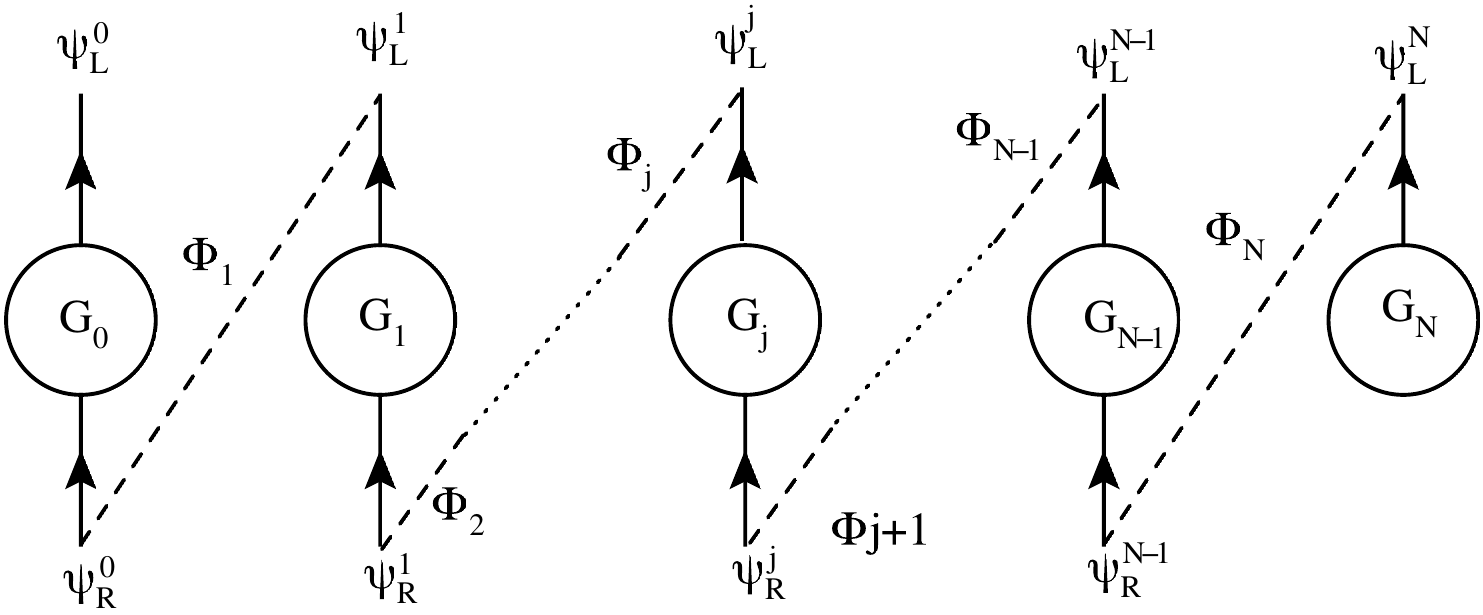}
\caption{Quiver diagram for the theory described by Eq.~(\ref{sf}),
  with a left-handed zero mode fermion.}
\label{fig:fermions}
\end{center}
\end{figure}

The introduction of fermions in these models  was extensively studied
in Refs.~\cite{bbh,quiver1}.  The fermion action is given by
\bear
S_f& =& \int d^4x \sum_{j=0}^N \left\{ \bar\psi_L^{j} i\hspace*{-0.1cm}\not\hspace*{-0.1cm}D_j \psi_L^{j}
+ \bar\psi_R^{j} i\hspace*{-0.1cm}\not\hspace*{-0.1cm}D_j \psi_R^{j} \right.\\
& &
\left. - (\mu_j \bar\psi_L^{j}\psi_R^{j} +
\lambda_j\bar\psi_R^{j-1}\Phi_j\psi_L^{j} + {\rm h.c.} )
\right\}~,
\label{sf}
\eear 
where the $\mu_j$ are vector-like masses and the Yukawa couplings are
chosen in such a way so as to only result in one zero mode
fermion~\cite{bbh}. 
For a left-handed zero-mode, the ``boundary condition'' must be chosen
such that $\psi_R^N=0$. Conversely, to obtain a right-handed zero mode
fermion, the condition is $\psi_L^0=0$. A schematic diagram of the
fermionic action is shown in Figure~\ref{fig:fermions} for a
left-handed zero mode. 
By using Eq.~(\ref{phis}), we can obtain the fermion mass matrix, which
just as for the case of gauge bosons is not diagonal due to the mixing
induced by the VEVs of the link fields $\Phi_j$. The rotation to a
mass-eigenstate basis is defined as
\be
\psi_{L,R}^{j} = \sum_{n=0}^N h_{L,R}^{j,n} \,\chi_{L,R}^{(n)} ~,
\label{rotation}   
\ee 
where the $\chi_{L,R}^{(n)}$ are the mass eigenstates. We are
interested in the coefficients $h_{L,R}^{j,0}$ corresponding to the
zero-mode localization in the quiver diagram.  They can be chosen so
as to obtain the correct fermion masses and mixings considering that
the Higgs is highly localized close to the site $N$.  For instance,
the situation with the Higgs localized at site $N$ was studied in  
Ref.~\cite{quiver1} for the quark sector. From the equations of motion
it is possible to obtain relations among the zero-mode
coefficients. In general, the zero-mode coefficients
for the left and right handed cases satisfy:
\bear
\sqrt{2}\frac{\mu_j}{v \lambda_{j+1}} &=& - q^{j+1}\,\frac{h_L^{j+1,0}}{h_L^{j,0}}\label{zmf1}\\ 
\sqrt{2}\frac{\mu_j}{v \lambda_j} &=& -
q^j\,\frac{h_R^{j-1,0}}{h_R^{j,0}}~.
\label{zmf2}
\eear
 The choice of fermion
localization can then be parametrized in order to get the desired
ratios in Eqns.~(\ref{zmf1}) and (\ref{zmf2}). For instance, we choose
the parametrizations
\be
\frac{h_L^{j+1,0}}{h_L^{j,0}} = q^{c_L -1/2}~,\qquad 
\frac{h_R^{j,0}}{h_R^{j-1,0}} = q^{-(c_R + 1/2)}  ~,
\label{cdefs}
\ee
which in the continuum limit would result in fermion zero-mode wave
functions parametrized by  $c_L$ and $c_R$ 
defined in \cite{rs}. As mentioned above, the localization
parameters $c_L$ and $c_R$ are chosen so as to obtain the observed 
pattern of fermion masses and mixings for a given Higgs localization
model. This can be a simple   N-localized Higgs as in
Ref.~\cite{quiver1}, or the dynamically localized pNGB as in Ref.~\cite{quiver2}.  

In the next section we will obtain the couplings of zero-mode fermions
(the SM fermions) to the first excitation of gauge bosons so we can
study their phenomenology at the LHC. 
 
\section{Resonances in Quiver Theories}
\label{resonances}
We are interested in obtaining  the couplings of the massive gauge
boson resonances  to the SM fermions. We follow closely Ref.~\cite{quiver1}.
The couplings  are defined by 
\begin{equation}
g^{01}_{L,R}\, \bar\chi_{L,R}^{(0)}\gamma^\mu A_\mu^{(1)} \chi_{L,R}^{(0)}~,
\label{g01coup}
\end{equation}
where we assumed that the  group generators are absorbed in the definition
of the gauge fields. The wave-function of a zero-mode fermion can be written  in terms of the quiver fermions as
\begin{equation}
\chi^{(0)}_{L,R} = \sum_{j=0}^N\,h_{L,R}^{*j,0}\,\psi_{L,R}^j~.
\label{zmfermion}
\end{equation}
In the same way, and assuming a generic
gauge group in the sites of the quiver diagram, the mass-eigenstates
of the gauge bosons can be written in terms of the quiver gauge bosons
as
\begin{equation}
A_\mu^{(n)} = \sum_{j=0}^N\, f^*_{j,n}\,A^j_\mu~,
\label{nmode}
\end{equation}
with $f_{j,n}$ the coefficient linking the gauge boson in site $j$
with the mass-eigenstate $n$ in the rotation to mass eigenstates. Therefore, the coupling of the  $n=1$ massive gauge boson to the zero-mode fermions is
\be
g^{01}_{L,R} = \sum_{j=0}^N\,
g_j\,\left|h_{L,R}^{j,0}\right|^2\,f_{j,1}~,
\label{g01}
\ee
where $g_j$ are the gauge couplings associated to the group $G_j$  in
the quiver and as mentioned before, we assume $g_j=g$ for all
$j$ in the manner defined by Eq.~(\ref{gaugeequal}).
The coefficients $f_{j,1}$ can
be obtained by diagonalizing the gauge boson mass matrix~\cite{tools,bbh} for a given $N$. Then, we can
obtain the couplings of zero-mode fermions to the first excited state
of the gauge bosons, normalized by the gauge coupling $g$.

\begin{figure}[!h]
\begin{center}
\includegraphics[scale=0.70]{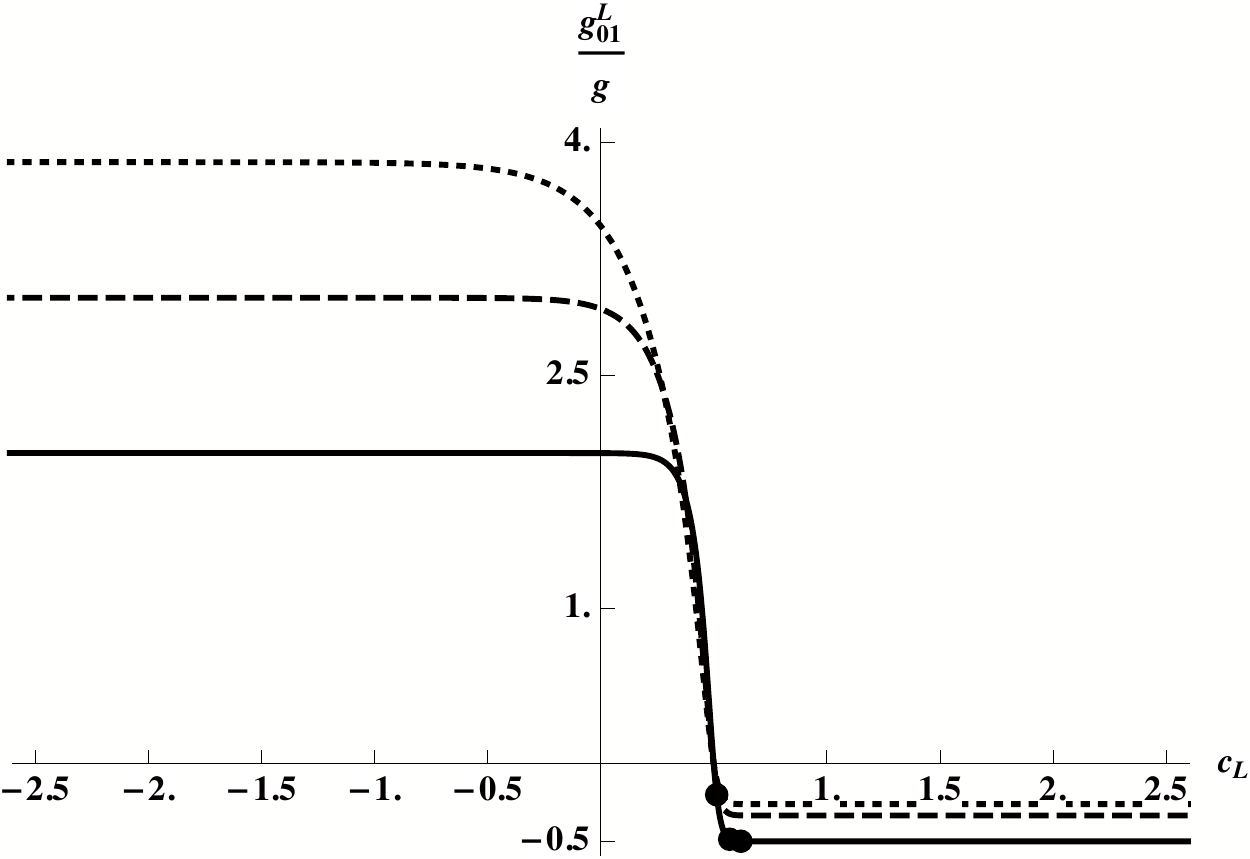}
\caption{The couplings of left-handed zero-mode fermions to the first
  excited gauge bosons as a 
function of the fermion localization parameter $c_L$,  for
$N=4~\mathrm{(solid)},$ $N=9~\mathrm{(dashed)},$ and
$N=15~\mathrm{(dotted)}$.
The dots correspond to the localizations for a solution for the $N=4$
case and are shown as an illustration.}
\label{fig:g01L}
\end{center}
\end{figure}

In Figure~\ref{fig:g01L} we show the couplings of the left-handed zero-mode
fermions to the first excited gauge boson state, $g^{L}_{01}$,
normalized to the zero-mode gauge coupling and for 
   $N=4$, $9$ and $15$, as a function of the fermion localization
   parameter  $c_L$ defined by Eqs.~(\ref{cdefs}). 

The values of the localization parameter above $c_L=0.5$ correspond to
``UV'' zero-mode localization:  most of the zero-mode wave function
comes from fermions transforming under gauge groups that are
associated with larger VEVs. Conversely, for $c_L< 0.5$ the zero-mode 
fermion wave-function is mostly coming from fermions transforming
under gauge groups associated with smaller VEVs. We refer to the
latter as ``IR'' localization. 

\begin{figure}[!h]
\begin{center}
\includegraphics[scale=0.70]{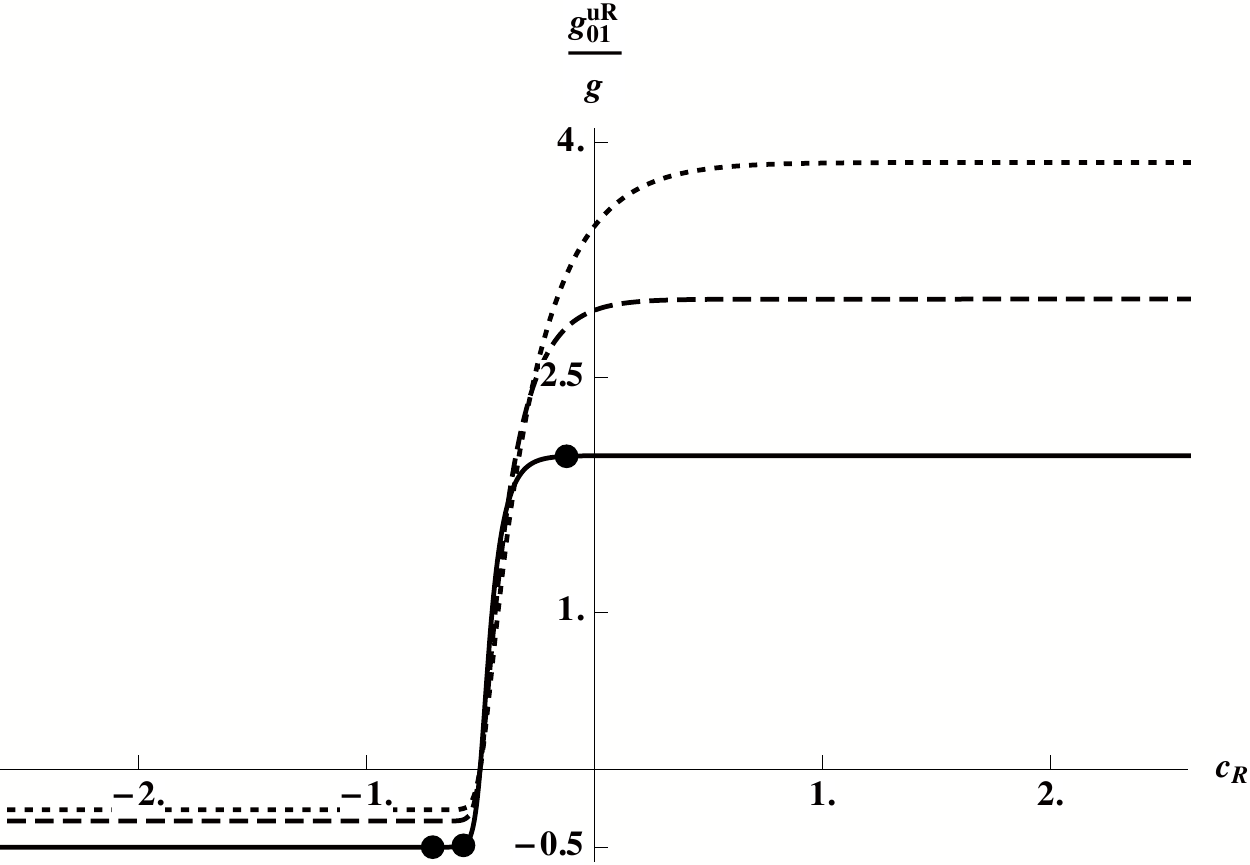}
\caption{The couplings of up-type right-handed zero-mode fermions to the first
  excited gauge bosons as a 
function of the fermion localization parameter $c_R$,  for $N=4~\mathrm{(solid)},$ $N=9~\mathrm{(dashed)},$ and $N=15~\mathrm{(dotted)}$.}
\label{fig:g01ur}
\end{center}
\end{figure}

\begin{figure}[!h]
\begin{center}
\includegraphics[scale=0.70]{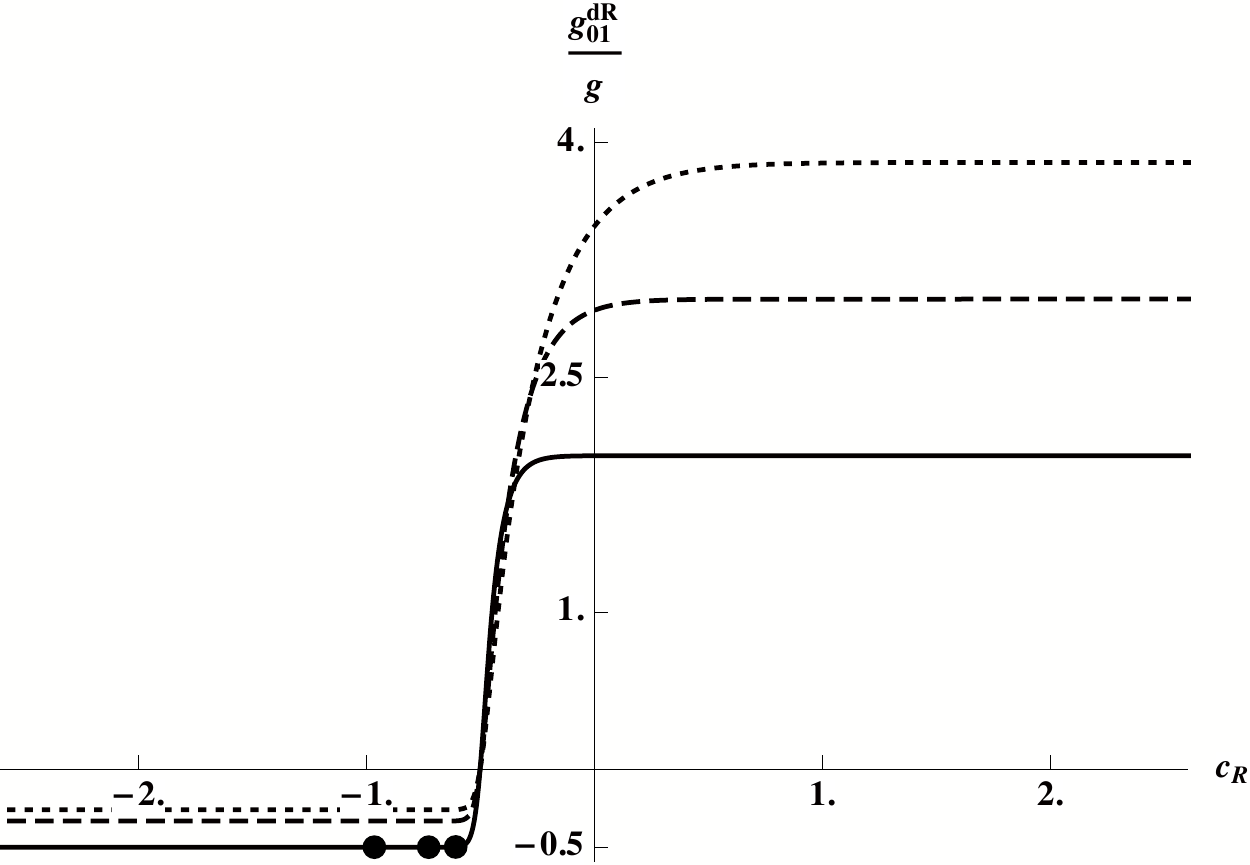}
\caption{The couplings of down-type right-handed zero-mode fermions to the first
  excited gauge bosons as a 
function of the fermion localization parameter $c_R$,  for $N=4~\mathrm{(solid)},$ $N=9~\mathrm{(dashed)},$ and $N=15~\mathrm{(dotted)}$.}
\label{fig:g01dr}
\end{center}
\end{figure}

Similarly, in Figures~\ref{fig:g01ur} and ~\ref{fig:g01dr} we show
the couplings of up and down right-handed zero-mode fermions to the first gauge
boson excitation, as a function of the respective $c_R$ localization
parameters. 
In these cases, localization parameters with values $c_R<-0.5$,
correspond to ``UV'' localization in the quiver, whereas for
$c_R>-0.5$, most of the zero-mode wave function comes from fermions
transforming under ``IR'' gauge groups.
The localizations illustrated by three points in
Figures~\ref{fig:g01L}, ~\ref{fig:g01ur} and ~\ref{fig:g01dr}
correspond to a given solution for the localizations of the zero-mode
quarks for $N=4$. This solution  is consistent with the quark mass
spectrum, the CKM matrix elements and has minimal  flavor-changing
neutral current effects~\cite{quiver1}. Similar solutions can be found
for the other values of $N$. 

As it can be seen in the Figures above, the couplings of IR-localized
zero-mode fermions increase with $N$, whereas the ones corresponding
to UV-localized, decrease. In the continuum limit, which as we noted
in the previous section is reached for $N\gae 36$, the couplings will
behave exactly as those in AdS$_5$ bulk models~\cite{rs}. However, and
as it was shown in Ref.~\cite{quiver1}, for coarse deconstruction
($N<36$), the resulting models have a different quantitative
behavior. For instance, flavor violation can be easily accommodated  
with mass scales above just a few TeV (specifically, $M_G>3~$TeV for
$N=4$ with $M_G$ the mass of the first excitation of the gluon),
whereas the continuum requires typically higher mass scales for the
Kaluza-Klein states. 

We can also see that the widths of the first excitations of gauge bosons
will not be as dominated by third generation channels as in the
continuum case. On one hand, the light UV-localized quarks leading to
jets have couplings to the excitation that are not as suppressed as in
the continuum case. Furthermore, the third generation couplings are
not as large.  In addition, the overall values of the couplings are
smaller, leading to significantly smaller total widths. Typical widths
for the first gauge excitations are $\Gamma/M\simeq 0.05$.  
These facts  result in a distinct phenomenology for resonance
production and decay when compared with the AdS$_5$ case.
For quiver theories, resonances will be narrow and
with significant di-jet signals. There will be still important
contributions to the $b\bar b$ and $t\bar t$ channels. The latter
might even dominate the bounds in some cases, as we will see below.  

In the next section, we  use the couplings computed here  
to obtain the s-channel production of the first-excited states of the
gauge bosons at the LHC into jets  and $t\bar t$ final states. 

\section{Resonances from Quiver Theories at the LHC}
\label{bounds}
In this section we study  the production of the first excited state
of the gauge bosons from full-hierarchy quiver theories at the LHC.  
We will consider two cases of particular interest. 

The first case,
corresponds to the quiver gauge group
with ${\cal G} = SU(3)^{N+1}$, broken down to the QCD gauge group, ${\cal G}\to
SU(3)_c$. The zero-mode gauge boson is the SM gluon, and the tower of
excited states are massive color-octet spin-1 resonances. 
This can be seen as the coarse deconstruction of bulk QCD in AdS$_5$. 
The reason to study this case is partly phenomenological: since they are
color-octet states they will have larger production cross
sections. It also  serves as comparison with the 
extra-dimensional case in  AdS$_5$
models with bulk $SU(3)_c$ gauge fields. 
However, unlike in the AdS$_5$ case, it is not necessary for $SU(3)_c$
to ``propagate'' in the quiver. It is entirely possible to obtain a
quiver model of EWSB and fermion masses with a pNGB Higgs boson
without a color-octet tower.  

The second case corresponds to having the quiver gauge group ${\cal G} =
(SU(2)_L\times U(1)_Y)^{N+1}$, broken to the electroweak SM gauge
group: ${\cal G}\to SU(2)_L\times U(1)_Y$. The zero-mode gauge bosons
in this case are the electroweak gauge bosons before EWSB, with them
replicated in the tower of excited states. 
The interest in this second
case resides on the fact that, although  in quiver models where the Higgs is a
pNGB~\cite{quiver2}  the quiver gauge groups must be
larger than the SM gauge group in order to extract the Higgs from
un-eaten NGBs,   
the massive states will contain these ones as a
subset. Thus, studying the phenomenology of these massive states is independent of the particular model chosen for the
electroweak quiver.

We will compute the cross section for production and decay to a given
channel for the color-octet and electroweak first vector resonances
at the LHC with $\sqrt{s}=8~$TeV, for various values of the number of
sites $N$. We concentrate on channels with quarks in the final state,
leading to light jets and $t\bar t$ final states. 
We leave out for now $b\bar b$ final states since there will be less
constraining. 
In each case the couplings to the SM quarks, the
zero-mode quarks  in the model as presented above, is computed
assuming a quark localization in the quiver consistent with the
correct mass matrix and  CKM mixing. These solutions for each value of
$N$ are then consistent with all flavor phenomenology.

\begin{figure}[!h]
\begin{center}
\includegraphics[scale=0.70]{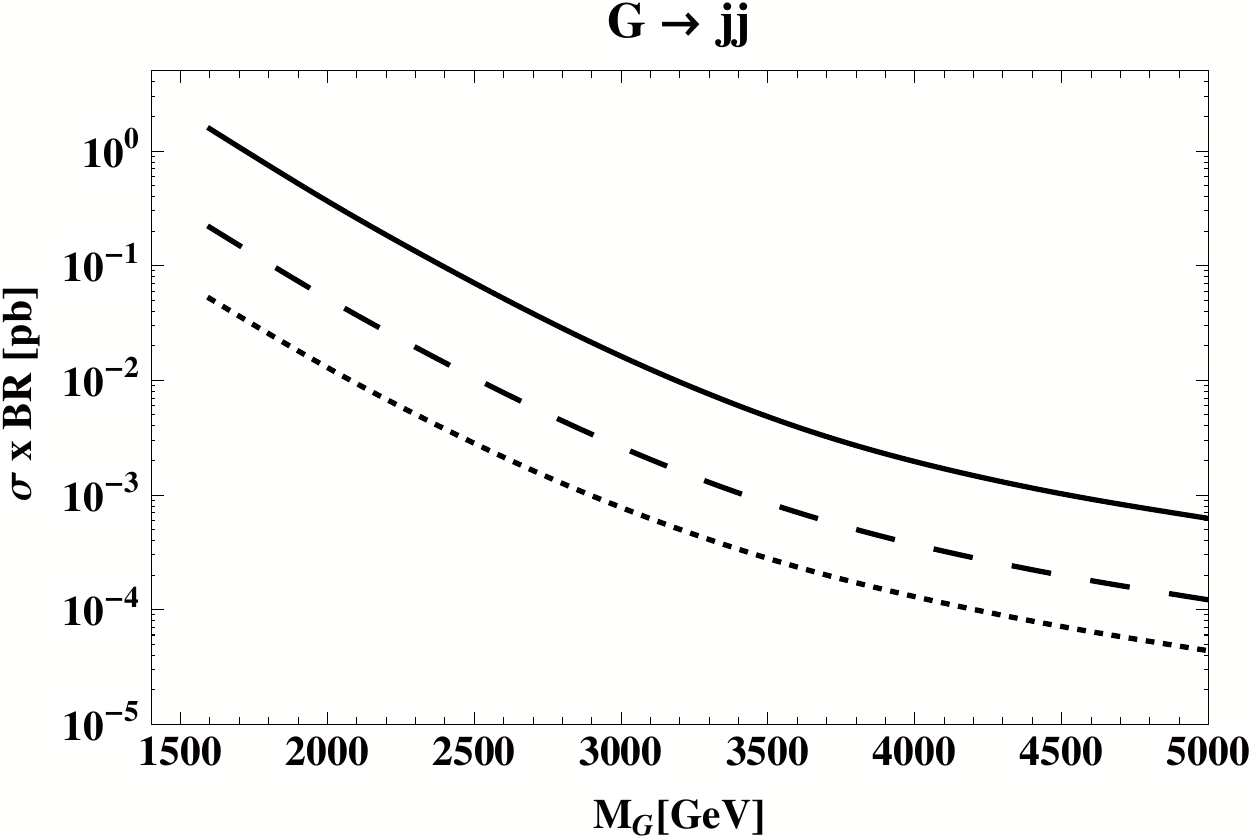}
\caption{The dijet production cross section times branching ratio for
  $\sqrt{s}=8~$TeV, as a
function of the  color-octet  mass for $N=4~\mathrm{(solid)},$ $N=9~\mathrm{(dashed)},$ and $N=15~\mathrm{(dotted)}$.}
\label{fig:gjj}
\end{center}
\end{figure}

\begin{figure}[!h]
\begin{center}
\includegraphics[scale=0.70]{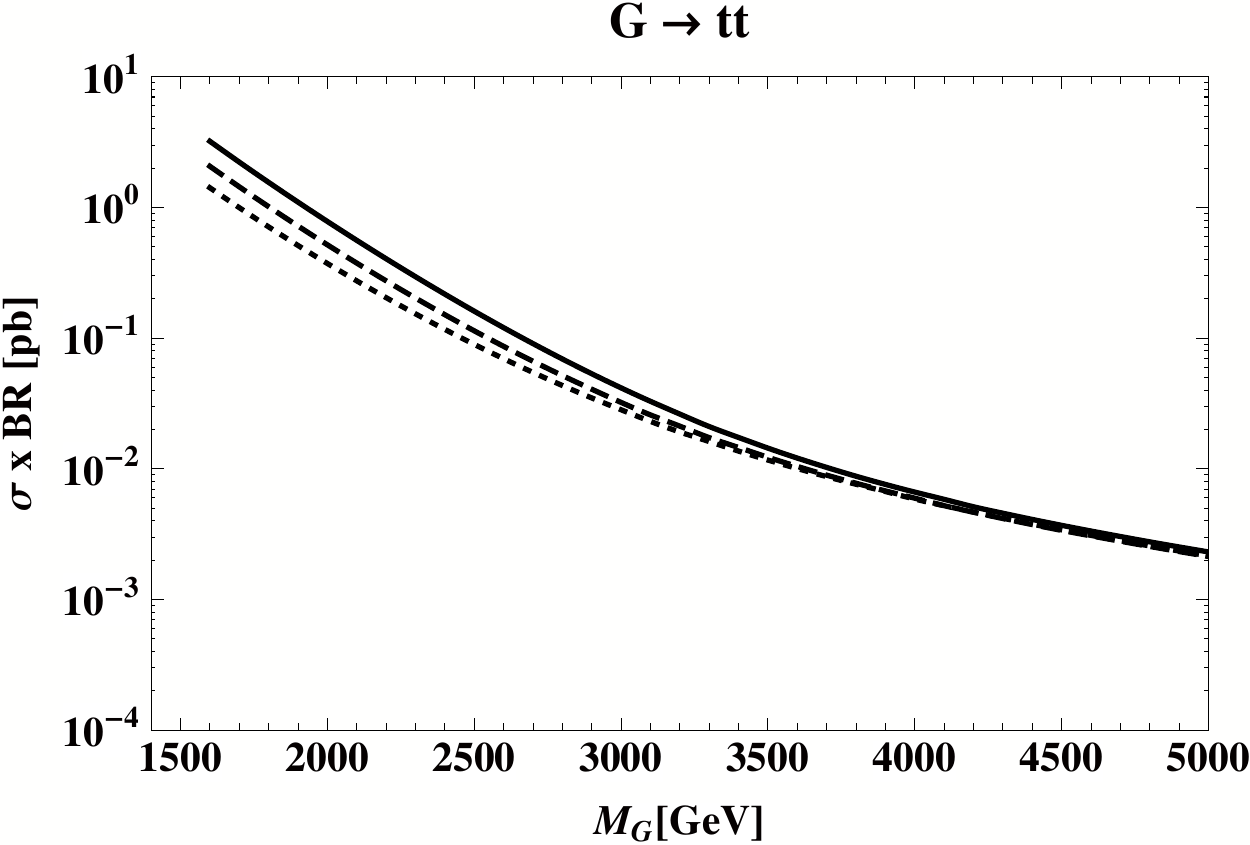}
\caption{The $t\bar t$ production cross section  times branching ratio
  for $\sqrt{s}=8~$TeV,  as a
function of the  color-octet  mass for $N=4~\mathrm{(solid)},$ $N=9~\mathrm{(dashed)},$ and $N=15~\mathrm{(dotted)}$.}
\label{fig:gtt}
\end{center}
\end{figure}

The  resonance widths are quite small in all cases. This is to be
compared to the AdS$_5$ situation where typical widths for the
Kaluza-Klein gluon are well above the typical resolution~\cite{kkgwidth}. 
We start with the color-octet excited states.The production cross section times the branching ratio into jets is
shown in
Figure~\ref{fig:gjj}, for three choices of the number of gauge groups
in the quiver: $N=4$ (5 gauge groups), $N=9$ and $N=15$. 
The corresponding plots for the color-octet production decaying into
a $t\bar t$ pairs is shown in Figure~\ref{fig:gtt}. 

\begin{figure}[!h]
\begin{center}
\includegraphics[scale=0.70]{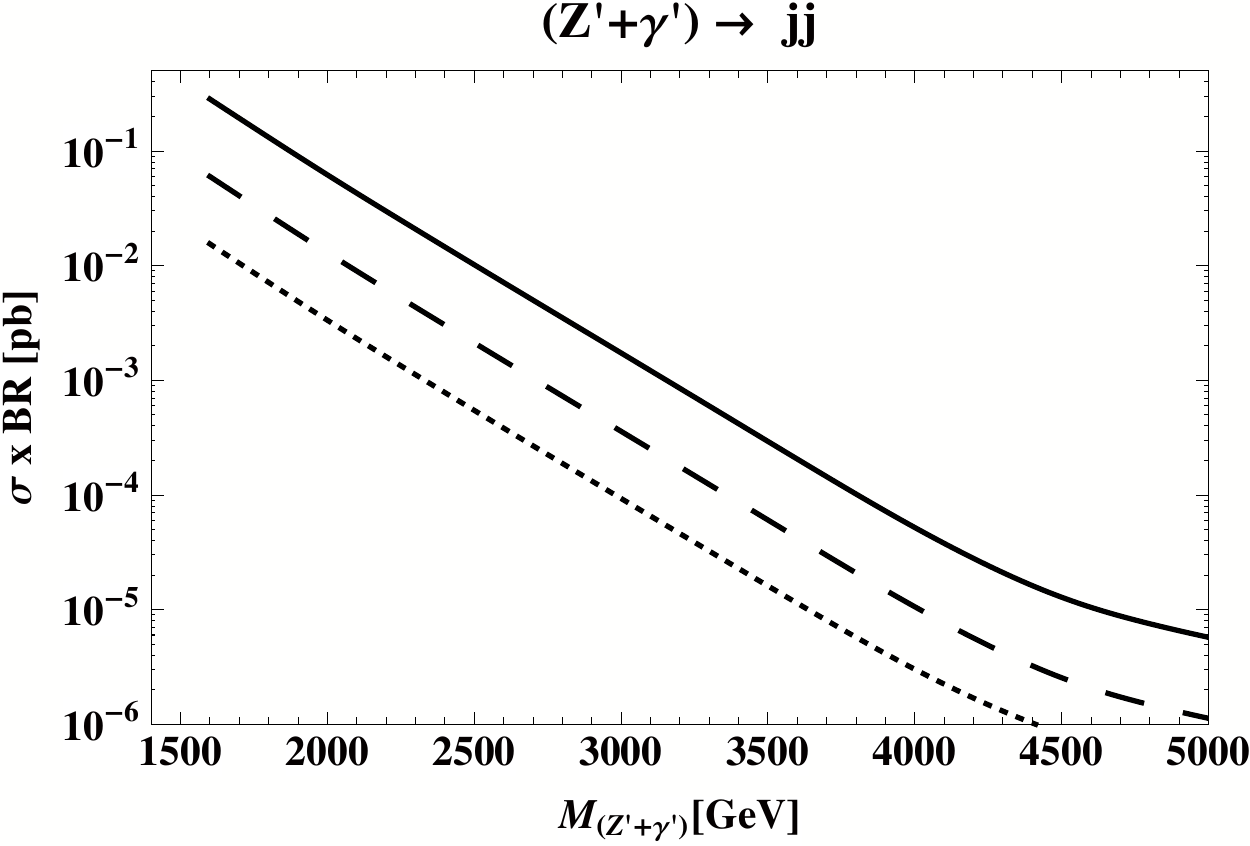}
\caption{The di-jet  production cross section  times branching ratio
  for $\sqrt{s}=8~$TeV,  as a
function of the $(Z'+\gamma')$ mass for $N=4~\mathrm{(solid)},$ $N=9~\mathrm{(dashed)},$ and $N=15~\mathrm{(dotted)}$.}
\label{fig:zpjj}
\end{center}
\end{figure}

\begin{figure}[!h]
\begin{center}
\includegraphics[scale=0.70]{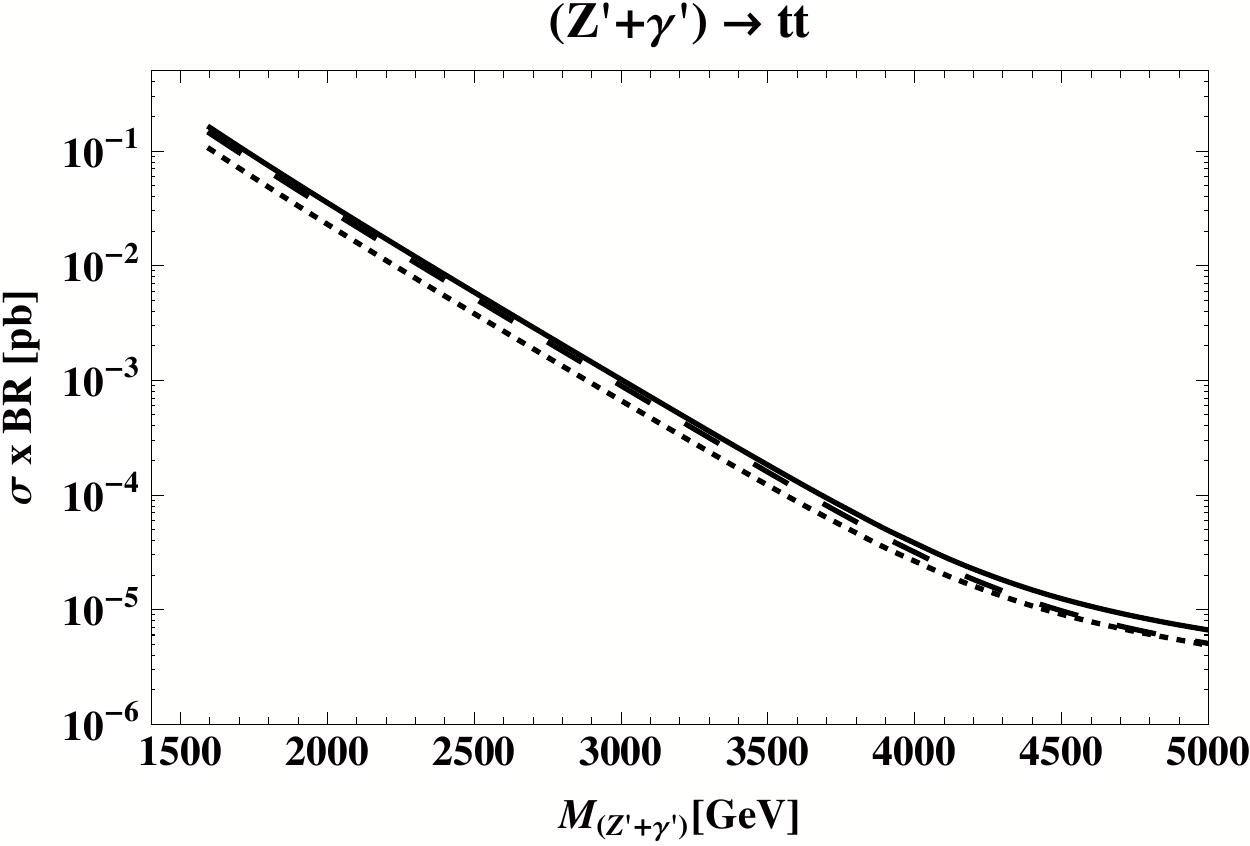}
\caption{The $ t\overline{t}$ production cross section times branching
  ratio $\sqrt{s}=8~$TeV, as a
function of the $(Z '+\gamma')$ mass for $N=4~\mathrm{(solid)},$ $N=9~\mathrm{(dashed)},$ and $N=15~\mathrm{(dotted)}$.}
\label{fig:zptt}
\end{center}
\end{figure}

We also consider color-singlet states, as mentioned earlier, as a
combination of the first excitation of the photon and the $Z$,
$(Z'+\gamma')$, since these are likely to be close in mass. 
In Figure~\ref{fig:zpjj}, we show the production times branching
ratios for $(Z'+\gamma')$ decaying to di-jets for several choices
of  $N$.  
A similar plot for the decays of the color-singlet into top pairs is
shown in Figure~\ref{fig:zptt}.

In the case of the di-jet decay channels, both for the color-octet as
well as for the color-singlet,  we can see that as the number of sites
in the quiver diagram ($N$)  grows, the $\sigma\times \rm{BR}$ falls
(Figures~\ref{fig:gjj} and \ref{fig:zpjj}).  This is to be expected, since
as $N$ grows and we approach the continuum AdS$_5$ limit the size of
the light quark couplings to the first gauge excitation diminishes.
On the other hand, the corresponding $t\bar t$ decay channels are
much more degenerate, as it can be seen in Figures~\ref{fig:gtt} and
\ref{fig:zptt}.  This is due to the fact that the top couplings to the
first gauge excitation grow with $N$, which almost exactly
compensates the reduction in production cross section coming from
smaller light-quark couplings.

We derive bounds from the LHC data accumulated with
$\sqrt{s}=8~\mathrm{TeV}$. In particular, we use the CMS bounds on
di-jets resonances of Ref.~\cite{CMSjj}, which uses
$19.6~\mathrm{fb}^{-1}$  of integrated luminosity, whereas we use the
bounds obtained by ATLAS on $t\bar t$ resonances~\cite{Atlastt} with 
an integrated
luminosity of 
$14.5~\mathrm{fb}^{-1}$.
Since the quiver resonances are narrow  interference effects can be
neglected. Moreover, in order to compare with  the experimental limits
we must only consider the resonance region since the bounds are
obtained by ``bump searches''. 
Table \ref{tab:co} shows the direct bounds from LHC on the
color-octet mass. These are obtained from the CMS constraints on
di-jet resonances in Ref.~\cite{CMSjj}, and from the ATLAS bounds on
$t\bar t$ resonances of Ref.~\cite{Atlastt}. 
\begin{table}[!h]
\begin{center}
\caption {Bounds on the color-octet mass [TeV].}
\begin{tabular}{|c|c|c|}
\hline\hline
$N$ & Dijet \cite{CMSjj} &  $t\overline{t}$ \cite{Atlastt} \\
\hline \hline
4 & 3.0 & 2.7 \\
\hline
9 & 1.6 &  2.6\\
\hline
15 & -- & 2.5\\ 
\hline
\end{tabular}
\label{tab:co}
\end{center}
\end{table}
We see that, unlike for the KK gluon in AdS$_5$ models, the di-jet
bounds are competitive, being the best limit in the $N=4$ case. As
mentioned above, the bounds coming from $t\bar t$ are not really
sensitive to $N$. All of the bounds on quiver resonances from
Table~\ref{tab:co} are similar to the flavor and electroweak precision
bounds obtained in Ref.~\cite{quiver1}, which were typically $\sim 3$~TeV.

We also consider the bounds on first electroweak gauge boson
excitations. As mentioned above, this sector typically contains at
least an
excitation of the $Z$ ($Z'$) and one of the photon ($\gamma'$), in
addition to other weakly coupled first excitations not
corresponding to any SM zero mode.  Here we study the bounds on this
minimum electroweak set of excitations, $Z'$ and
$\gamma'$. Furthermore, we will assume that their masses are close
enough to appear degenerate at the LHC, at least in the search
stages. As a consequence, we will obtain bounds on the $Z'+\gamma'$ combination.
\begin{table}[!h]
\begin{center}
\caption { Bounds on the ($Z'$+$\gamma '$) mass [TeV]}
\begin{tabular}{|c|c|c|}
\hline\hline
$N$  & Dijet \cite{CMSjj} &  $t\overline{t}$ \cite{Atlastt} \\
\hline \hline
4 & 1.7 & 2.1 \\
\hline
9 & -- &  2.0 \\
\hline
15 & --&  1.8 \\
\hline 
\end{tabular}
\label{tab:zp}
\end{center}
\end{table}
In Table~\ref{tab:zp} we show the bounds on the $(Z'+ \gamma')$
combination from di-jets from CMS~\cite{CMSjj}, and from top pairs
from ATLAS~\cite{Atlastt}. Once again, the di-jet channel is
competitive for low values of the number of sites, but  $t\bar t$ is
most constraining in general.  The entries without a bound, both in
Table~\ref{tab:zp} as well as in Table~\ref{tab:co}, correspond to
bounds that are too low for them to be consistent with flavor and
electroweak limits, as well as other direct bounds. 

We observe that the bounds obtained in Tables~\ref{tab:co} and
\ref{tab:zp} are still below what is needed to pass flavor-violation
bounds in most of these models ($\simeq 3$~TeV). This mass range will
be probed by the next stage of the LHC, with higher energy and
luminosity.

\section{Outlook and Conclusions}
\label{conclusions}
We have considered the phenomenology of a class of four-dimensional
quiver theories~\cite{quiver1}, related to AdS$_5$ bulk
models~\cite{rs}  by
coarse deconstruction. In particular, we have studied the current
bounds on gauge excitations in these theories imposed by using the
current LHC data. 
To be as general as possible we considered two kinds of
resonances. First, we studied a
color-octet excitation which corresponds to the propagation of color
$SU(3)$ in the quiver. Unlike in the extra-dimensional
formulation, this propagation is not necessary. However,  we consider
this case for completeness and comparison to the AdS$_5$ case. 
Secondly, we studied the bounds on the minimal
electroweak excitations of these models, namely a $Z$ and photon
excitations. For simplicity, we assumed that these two are nearly
degenerate, whereas their common mass need not be the same as that of
the color-octet state. The rationale for this split in the spectrum is
that corrections to the color-octet mass should in general be
different and probably larger than the ones affecting  the colorless
states. This allows for the possibility that the color-octet state,
which drives the flavor bounds~\cite{quiver1}, is heavier than the 
electroweak excitations.

We treated the number of sites $N$ as a free parameter, as long as it
satisfies coarseness, i.e. $N \le 36$. In this way, the phenomenology
of these spin-1 resonances is guaranteed to be qualitatively
different from that of AdS$_5$ Kaluza-Klein states.  The bounds
obtained for the color-octet state and the weakly coupled combination
$(Z'+ \gamma')$ depend on the parameter  $N$, and they appear in
Tables~\ref{tab:co} and \ref{tab:zp}. We see that in both cases the
$t\bar t$ constraints are still consistently dominant. However, the
di-jet bounds can be competitive for lower number of sites, for which
the light quark couplings are not as suppressed. For instance, for
$N=4$ ($5$ sites), the most stringent bound comes from the di-jet
channel of the color-octet. Still in this case, we see that the bounds
are not yet above the mass scale needed to suppressed flavor
changing neutral current, typically $\gae 3$~TeV~\cite{quiver1}. 

For the colorless states the bounds obtained are somewhat smaller, as
shown in Table~\ref{tab:zp}. Although in principle these bounds are
consistent with flavor violation in the quark sector, the most
important constraints on these states will probably come from flavor
violation in the lepton sector. However, these
are not yet available for quiver theories, as their lepton sector is
only now beginning to be considered in the literature~\cite{leptons}. 

\begin{figure}[!h]
\begin{center}
\includegraphics[scale=0.70]{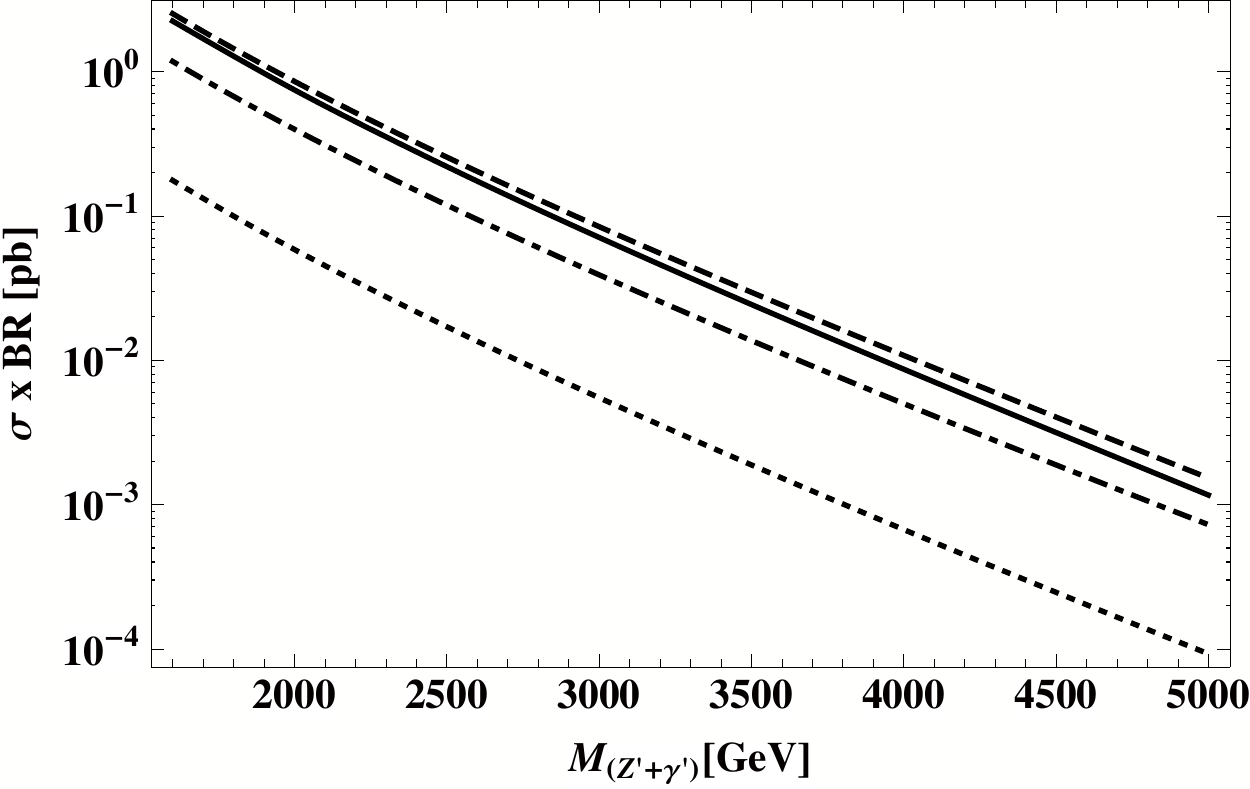}
\caption{The color-singlet production cross section times branching ratio as a
function of the  color-singlet  mass into di-jets for
 $N=4~\mathrm{(solid)},$ and  $N=15~\mathrm{(dotted)}$;  and into top
 pairs for $N=4~\mathrm{(dashed)}$ and
 $N=15~$(dot-dashed). All for $\sqrt{s}=14~$TeV. }
\label{fig:zp14}
\end{center}
\end{figure}

In order for direct searches to compete with the flavor bounds of Ref.~\cite{quiver1}, it
would be necessary to probe above the  
mass scale of about $3~$TeV. We conclude that to do this decisively,
the higher energy run at the LHC will be necessary. To illustrate this
point we show the cross sections for the production of the color-octet
and color-singlet states studied in this paper, at $\sqrt{s}=14~$TeV,
for the di-jet and $t\bar t$ channels. 
\begin{figure}[!h]
\begin{center}
\includegraphics[scale=0.70]{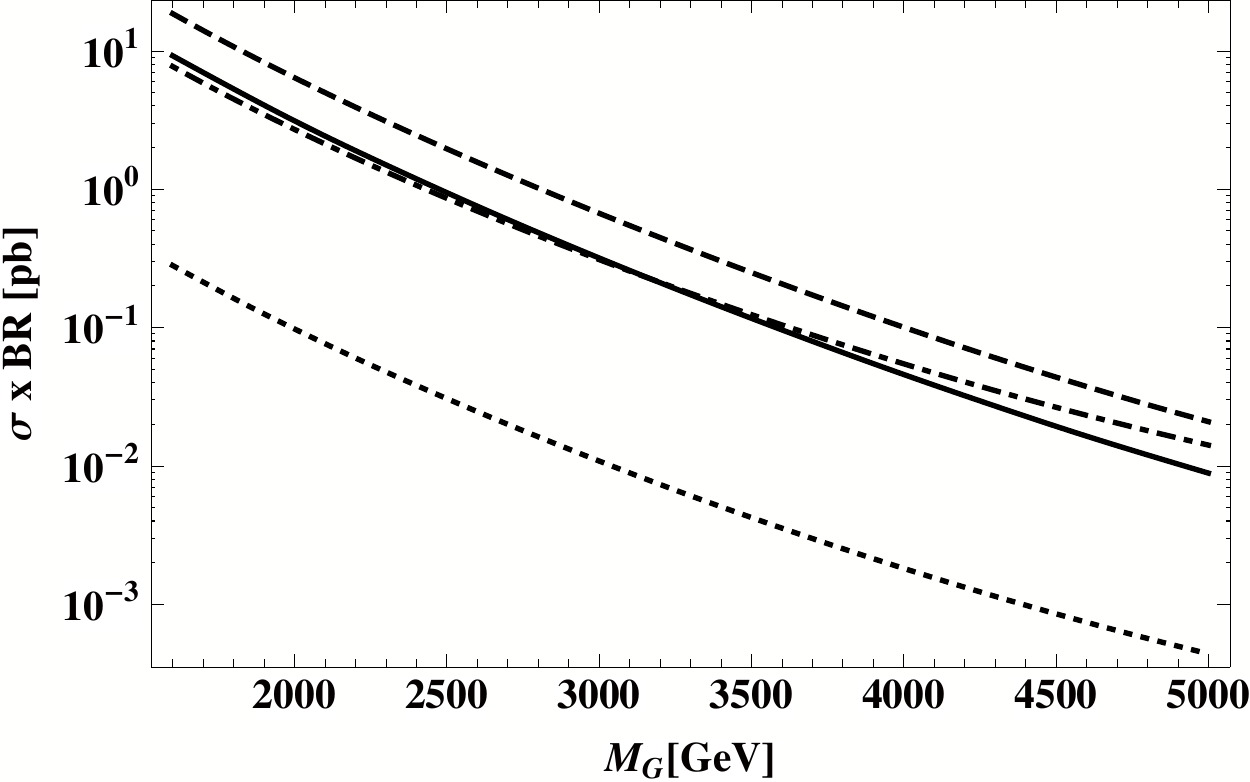}
\caption{The color-octet production cross section times branching ratio as a
function of the  color-octet  mass into di-jets for
 $N=4~\mathrm{(solid)},$ and  $N=15~\mathrm{(dotted)}$;  and into top
 pairs for $N=4~\mathrm{(dashed)}$ and
 $N=15~$(dot-dashed). All for $\sqrt{s}=14~$TeV. }
\label{fig:g14}
\end{center}
\end{figure}
In Figure~\ref{fig:g14} we show the color-octet production cross
sections times branching fractions into di-jets for $N=4$ (solid) and
$N=15$ (dotted), as well as the ones into $t\bar t$ for $N=4$ (dashed)
and $N=15$ (dot-dashed). Although a careful study is necessary to know
the reach of the LHC at $\sqrt{s}=14~$TeV for a given luminosity, we
can see that the reach in the color-octet mass will be much above
3~TeV, perhaps as much as 5~TeV with a few hundred $\rm{fb}^{-1}$ of
accumulated luminosity. Similarly, cross sections times branching fractions for the electroweak states $(Z'+ \gamma')$,
for $\sqrt{s}=14~$TeV are shown in Figure~\ref{fig:zp14}.

We have seen that the quiver theories studied here are phenomenologically
distinct from AdS$_5$ models. In particular, the existence of rather
narrow resonances even in the color-octet case would point to states  very
different from a Kaluza-Klein gluon. Quiver theories generalize the
model building philosophy of AdS$_5$ models of electroweak symmetry
breaking and fermions masses. The spin-1 resonances studied here
should be among the first signals for these kind of physics. Other
signals, parametrized by the number of sites $N$, would follow. Their
study would  depend on details of the models,  such as fermion
representations chosen, the model building of the lepton
sector~\cite{leptons} and the Higgs sector~\cite{quiver2}, just to
mention a few. Ultimately, quiver  theories form a class of theories
beyond the SM which includes AdS$_5$ as the continuum limit. Thus, their
phenomenology at the  LHC  
should be treated together. For instance, the presence of a set of signals
for new physics could  determine the value of $N$ (if any) consistent 
with all of them.  The theoretical interpretation of this value, whether
indicating a continuum theory or a coarse quiver one, would be an
important step in determining the road to build the right theory of
the TeV scale.

\vskip0.3in 
\noindent{\bf Acknowledgments }
The authors  acknowledge the support of the State of S\~{a}o Paulo
Research Foundation (FAPESP),  the Brazilian  National Council
for Technological and Scientific Development (CNPq), and the Brazilian
Agency for Postgraduate Development (CAPES).

\end{document}